# Constructive Port-Hamiltonian Energy Shaping Design of Dispatchable Virtual Oscillators in Grid-Forming Converters

Lu Gao, Lihui Yang, Feng Ji, Dong Liu, Longze Kou

***Abstract*—Port-Hamiltonian (PH) theory offers a passivity-based framework for grid-forming control, yet conventional dispatchable virtual oscillator control (dVOC) does not naturally admit a dissipative PH realization, since its amplitude regulation, synchronization, and power dispatch are inherently coupled without a unified energy interpretation. This paper formulates the outer-loop dynamics as a dissipative PH system, thereby unifying amplitude regulation, synchronization, and power dispatch within one energy structure. The formulation rests on the key property that logistic-type radial regulation, characterized by an inherent saturation-like nonlinearity, permits an exact gradient decomposition compatible with the quadratic energy storage. On this basis, a unified shaped Hamiltonian is constructed, which encapsulates both amplitude restoration and power dispatch. Radial gain matching derived from this Hamiltonian yields explicit closed-form parameter inequalities that guarantee almost-global asymptotic stability and local exponential convergence. Moreover, tuning the power-error weighting coefficient actively shapes the energy landscape, thereby eliminating the undesirable low-voltage power-flow solution from the stationary set and ensuring convergence to the desired high-voltage equilibrium point. The Hessian singularity condition further provides the critical weight threshold that guarantees uniqueness of the high-voltage equilibrium. Numerical simulations validate the proposed method.**

***Index Terms*—energy shaping, grid-forming converters, Logistic-type amplitude regulation, port-Hamiltonian system, virtual oscillator control.**

## I. Introduction

As the penetration of renewable energy sources in power systems continues to increase, power converters are progressively required to undertake functionalities such as voltage establishment, frequency support, and power regulation. Grid-forming control, by actively generating voltage amplitude and phase at the converter terminals, enables converters to maintain AC system operation even in the absence of strong synchronous sources, and has thus become a critical control paradigm in high-penetration power-electronics-dominated systems. Virtual oscillator control (VOC) employs nonlinear oscillators to generate stable limit cycles, allowing parallel-connected converters to achieve self-synchronization and power sharing based solely on local electrical quantities [1]. So it has become an important outer loop control structure for grid forming converters. Building upon this, dispatchable virtual oscillator control directly incorporates active and reactive power references into the voltage vector dynamics, endowing converters with both oscillator synchronization characteristics and power dispatch capability, and establishing a stability analysis framework for phase and amplitude synchronization [2]-[4].

Existing research on dVOC has yielded substantial results. Early work primarily relied on linearized models to analyze local stability around given operating points. In recent years, research focus has gradually shifted toward nonlinear and large-disturbance problems [5]-[8]. However, the design of dVOC injects power tracking errors directly into the oscillator dynamics, and the coupling among amplitude self-excitation, phase synchronization, and power dispatch lacks a unified analytical energy structure. This "functional splicing" design approach renders the nonlinear mechanisms of the controller difficult to systematically interpret, and limits its extension to complex scenarios such as multi-converter interconnections and nonlinear stability under large disturbances. To this end, a theoretical framework is needed that employs energy as a unified language, incorporating physical dissipation, power flow, and closed-loop stability into a single geometric structure.

Port-Hamiltonian system theory provides precisely such a unified framework. Through the explicit separation of interconnection structures and dissipation structures, it provides a modular description of energy storage, exchange, and dissipation in physical systems, and is naturally compatible with passivity-based analysis and energy shaping design. In recent years, PH methods have been introduced into the stabilization design of grid-forming converters, with a series of advances achieved[9]-[15]. Nevertheless, existing works primarily target stabilization around given equilibrium points or achieving closed-loop passivity, and have not fully exploited the energy shaping capability of the PH framework to uniformly handle amplitude restoration and power dispatch within the energy function. Meanwhile, there exists a structural mismatch between the self-sustained oscillator mechanisms in conventional VOC based on Van der Pol-type limit cycles [1] and the typical quadratic energy storage structures in the PH framework. A systematic energy-shaping design that incorporates the core functionalities of dVOC into a standard

Lu Gao, Lihui Yang, are with School of Electrical Engineering Xi'an Jiaotong University, Xi'an, Shaanxi 710049, China (e-mail:gl_xjtu@stu.xjtu.edu.cn, lihui_yang@mail.xjtu.edu.cn).

Feng Ji, Dong Liu, Longze Kou are with China Electric Power Research Institute Co., Ltd., Beijing 100192, China(jameskeating@163.com, 16039662@qq.com, koulongze@163.com.

PH form is therefore still lacking. Moreover, existing dVOC stability results, such as [2], typically require a posteriori verification of coupled gain--network conditions. This limits their direct use for controller synthesis with prescribed dynamic specifications. The proposed PH formulation instead yields explicit closed-form design inequalities that relate the prescribed convergence rate and outer-loop bandwidth to independently selectable controller gains. Accordingly, the outer-loop structure should be redesigned to retain the power-dispatch and synchronization capabilities of dVOC while enabling constructive energy shaping with a quadratic Hamiltonian within the PH framework.

To this end, this paper starts from the energy function rather than augmenting an oscillator with dispatch functionalities: a Logistic-VOC outer loop is formulated from the perspective of port-Hamiltonian energy shaping, and a complete energy-shaping design is developed on the resulting dissipative port-Hamiltonian model. The main contributions of this paper are as follows: (1)A dissipative port-Hamiltonian outer-loop model is established with the port voltage vector as the state, and a shaped Hamiltonian is constructed to unify amplitude restoration and power dispatch within a single energy structure. (2)The logistic-type radial regulation is revealed to admit an exact decomposition into a state-dependent gain and a quadratic potential-energy gradient, enabling its embedding into a dissipative PH model. (3)A state-dependent dissipation matrix is designed to achieve radial gain matching, guaranteeing closed-loop dissipativity, almost-global asymptotic stability, and local exponential convergence. Explicit closed-form inequalities are derived for direct parameter tuning. (4)The power-error weighting coefficient is shown to shape the energy landscape, eliminating the low-voltage solution from the stationary set and ensuring convergence to the high-voltage equilibrium. The critical weight threshold is explicitly determined.

## II. dVOC Baseline and Grid-Connected Outer-Loop Model

### A. Dispatchable Virtual Oscillator Control

The dispatchable virtual oscillator control proposed in [2] generates the voltage reference for grid-forming converters through three additive terms: a rotational term, a network-coupling term, and an amplitude-regulation term. The control law in the stationary reference frame is given by

$$\dot{v}_{\alpha\beta} = \omega_0 J v_{\alpha\beta} + \eta_d (K_d v_{\alpha\beta} - y_d) + \alpha_d \Phi_d (v_{\alpha\beta}) v_{\alpha\beta} \tag{1}$$

where $v_{\alpha\beta} = \begin{bmatrix} v_\alpha & v_\beta \end{bmatrix}^{\mathrm{T}}$ is the port voltage vector, $J = \begin{bmatrix} 0 & -1 \\ 1 & 0 \end{bmatrix}$ is the $90^\circ$ rotation matrix, $\omega_0$ is the rated angular frequency, and $\eta_d > 0$, $\alpha_d > 0$ are the synchronization gain and amplitude-regulation gain, respectively. Here, $K_d$ is constructed from the power and voltage set-points, and $y_d$ is a network feedback signal obtained from local current measurements (see [2, Eqs. (7) and (23)] for their explicit constructions). The amplitude-regulation function is defined as

$$\Phi_d = \frac{\rho_{ref} - \left\| v_{\alpha\beta} \right\|}{\rho_{ref}} \tag{2}$$

which drives the voltage magnitude toward the reference $\rho_{ref}$.

The three terms in (1) correspond to three distinct control objectives: $\omega_0 J v_{\alpha\beta}$ : imposes the rated frequency; $\eta_d (K_d v_{\alpha\beta} - y_d)$ : achieves phase synchronization and power dispatch; $\alpha_d \Phi_d (v_{\alpha\beta}) v_{\alpha\beta}$ : regulates the voltage magnitude.

Since the subsequent outer-loop model and power equations are formulated in the synchronous rotating frame, we transform the dVOC dynamics into the rotating frame following [2, Eq. (30)]. In this frame, the dVOC law reduces to

$$\dot{v} = \eta_d (K_d v - y_d) + \alpha_d \Phi_d (v) v \tag{3}$$

where $v = \begin{bmatrix} v_x & v_y \end{bmatrix}^{\mathrm{T}}$ denotes the port voltage vector in the synchronous rotating frame and $\Phi_d(v) = \dfrac{\rho_{ref} - \|v\|}{\rho_{ref}}$.

However, these functionalities are superimposed additively, and their coupling is not characterized through a unified energy function. This observation motivates the energy shaping design developed in Section III.

### B. Grid-Connected Outer-Loop Model

Consider a grid-forming converter connected to an ideal grid through a line inductance $L_g$. Let

$$v = \begin{bmatrix} v_x & v_y \end{bmatrix}^{\mathrm{T}}, v_g = \begin{bmatrix} v_{gx} & v_{gy} \end{bmatrix}^{\mathrm{T}} \tag{4}$$

denote the port voltage vector of the converter and the grid voltage vector, respectively, both expressed in the synchronous rotating frame. In this paper, we focus on the outer-loop voltage generation and power dispatch, assuming that the inner voltage/current loops can ideally track the voltage references generated by the outer loop.

The active and reactive powers injected by the converter into the grid in port voltage coordinates are

$$\begin{cases} P(v) = \dfrac{v^T J v_g}{\omega_0 L_g} \\ Q(v) = \dfrac{v^T v - v^T v_g}{\omega_0 L_g} \end{cases} \tag{5}$$

The state domain of the outer-loop is $D = \mathbb{R}^2 \setminus \{0\}$, since the polar-coordinate transformation is singular at the origin.

## III. Port-Hamiltonian Framework-Based Logistic-VOC Grid-Forming Control

This section completes the energy-shaping design: a shaped Hamiltonian function unifying amplitude establishment and power dispatch is constructed and its landscape analyzed, a radial gain-matching control law is then designed to enforce the port-dissipation inequality, regional stability is established, and a parameter design procedure is given.

### A. Logistic Radial Dynamics and Amplitude Potential Energy

We first characterize the Logistic-type radial voltage-restoration mechanism used in the proposed control law. In the synchronous rotating frame, the amplitude of the port voltage vector $v$ evolves as

$$\dot{\rho} = k\rho(1 - \frac{\rho}{\rho_{ref}}) \tag{6}$$

where $\rho = \|v\|$, $k$>0 is the radial regulation gain. Equivalently, the corresponding voltage-vector dynamics is

$$\dot{v} = k\left(1 - \frac{\rho}{\rho_{ref}}\right)v \tag{7}$$

Define the amplitude error as

$$\tilde{\rho} = \rho - \rho_{ref} \tag{8}$$

Define the amplitude potential energy function

$$H_1 = \frac{1}{2}\tilde{\rho}^2 \tag{9}$$

A direct computation gives the gradient

$$\nabla H_1 = \tilde{\rho}\frac{v}{\rho} \tag{10}$$

Substituting (10) into (7), the vector dynamics can be rewritten as

$$\dot{v} = -\frac{k\rho}{\rho_{ref}}\nabla H_1 \tag{11}$$

Equations (11) show that the Logistic-type radial regulation is governed by a state-dependent positive dissipation gain and the negative gradient of the amplitude potential energy. This property enables its incorporation into the port-Hamiltonian energy-shaping framework developed in the following subsection.

### B. Shaped Hamiltonian and Gradient Structure

This subsection constructs the shaped Hamiltonian that serves as the storage function for the port-Hamiltonian representation of the proposed outer-loop dynamics.

Define the energy coordinate $z(v) = \begin{bmatrix} \tilde{\rho} & \tilde{P} & \tilde{Q} \end{bmatrix}^T$, where $\tilde{P} = P - P_{ref}$, $\tilde{Q} = Q - Q_{ref}$. To jointly characterize voltage-magnitude regulation and power dispatch in a port-Hamiltonian framework, a positive definite weighting matrix $K = diag(k_\rho, k_P, k_Q)$ is configured, and the shaped Hamiltonian function is defined as

$$H(v) = \frac{1}{2}z^T K z \tag{12}$$

For convenience in subsequent control design, let $k_\rho = 1$, $k_P = k_Q = \alpha$, where α>0 is the weighting coefficient. The equivalent expression of (12) can be obtained as

$$H(v) = \frac{1}{2}\tilde{\rho}^2 + \frac{\alpha}{2}\tilde{P}^2 + \frac{\alpha}{2}\tilde{Q}^2 \tag{13}$$

Equation (13) explicitly decomposes the shaped Hamiltonian function into three weighted terms: amplitude potential energy, active power potential energy, and reactive power potential energy.

### C. Energy Gradient, Hessian Matrix, and Stationary Point Conditions

Taking the gradient of (13) with respect to the $v$ and applying the chain rule

$$\nabla H(v) = \left[\frac{\partial z(v)}{\partial v}\right]^T Kz(v) \tag{14}$$

The Jacobian matrix of the energy coordinates vector with respect to $v$ can be written from the power mapping

$$\frac{\partial z(v)}{\partial v} = \begin{bmatrix} \hat{v} & \frac{Jv_g}{\omega_0 L_g} & \frac{2v - v_g}{\omega_0 L_g} \end{bmatrix}^T \tag{15}$$

where $\hat{v} = \frac{v}{\rho}$.

Substituting $\hat{v}$ and $K = diag(1, \alpha, \alpha)$ into (14) and expanding, we get

$$\nabla H = \tilde{\rho}\hat{v} + \alpha\tilde{P}\nabla P + \alpha\tilde{Q}\nabla Q \tag{16}$$

where $\nabla P$ and $\nabla Q$ denote the gradients of $P$ and $Q$ with respect to $v$, respectively.

Using the power model, two radial identities can be obtained

$$v^T\nabla P = P, \; v^T\nabla Q = \frac{\rho^2}{\omega_0 L_g} + Q \tag{17}$$

The radial component of the power error gradient is further defined as:

$$C(v) = \alpha\tilde{P}P + \alpha\tilde{Q}(\frac{\rho^2}{\omega_0 L_g} + Q) \tag{18}$$

$$v^T\nabla H = \rho\tilde{\rho} + C(v) \tag{19}$$

Equation (19) indicates that the radial component of the shaped energy gradient contains not only the amplitude error $\tilde{\rho}$ but also $C(v)$ generated by the power errors. This coupling term is the reason why radial gain matching is required in the subsequent control law design.

For subsequent parameter design, the Hessian matrix of the energy function is also required.

The Hessian of the amplitude potential energy $\frac{1}{2}\tilde{\rho}^2$ with respect to $v$ can be computed as

$$A_\rho = (1 - \frac{\rho_h^*}{\rho})I + \frac{\rho_h^*}{\rho}\Pi_r \tag{20}$$

where $\Pi_r = \hat{v}\hat{v}^T$ denotes the radial projection matrix and $\rho_h^* = \rho_{\text{ref}}$ is the magnitude of the desired high-voltage equilibrium.

$$\nabla^2 H = A_\rho + \alpha\nabla P\nabla P^T + \alpha\left[\nabla Q\nabla Q^T + \frac{2\tilde{Q}}{\omega_0 L_g}I\right] \tag{21}$$

where $I = \begin{bmatrix} 1 & 0 \\ 0 & 1 \end{bmatrix}$.

### D. Radial Gain-Matching and Port-Hamiltonian Control Law

If gradient feedback is simply superimposed on the original Logistic amplitude term, one obtains

$$\dot{H} = -\frac{k\rho}{\rho_h^*}\tilde{\rho}^2 - \frac{k}{\rho_h^*}\tilde{\rho}C(v) - \eta\left\|\nabla H\right\|^2 \tag{22}$$

The cross term $\tilde{\rho}C(v)$ in $\dot{H}$ has indefinite sign; therefore, it cannot directly guarantee $\dot{H} \le 0$ . To make the radial component of the power error share the same gain as the Logistic amplitude channel, a radial correction term is added to the control law

$$\dot{v} = -\frac{k}{\rho_h^*}\tilde{\rho}v - \eta\nabla H - \frac{k}{\rho_h^*}C(v)\frac{v}{\rho} \tag{23}$$

Equation (23) is convenient for explaining the physical role of each control channel. The third term is the proposed radial correction term, which matches the radial gain of the power-error channel to that of the Logistic amplitude channel, thereby eliminating the sign-indefinite cross coupling in (19).

Substituting (18), (19) into (23), the control law can be rearranged as

$$\dot{v} = -(\eta I + \frac{k\rho}{\rho_h^*}\Pi_r)\nabla H \tag{24}$$

Equation (24) shows that the radial gain-matching correction modifies the gradient feedback through a state-dependent radial dissipation term. Define

$$\mathcal{J}(v) = 0,\ \ \mathcal{R}(v) = \eta I + \frac{k\rho}{\rho_h^*}\Pi_r \tag{25}$$

Since $\mathcal{J}(v) = -\mathcal{J}(v)$ , (24) can be written as

$$\dot{v} = [J(v) - R(v)]\nabla H(v) + g(v)u \tag{26}$$

Thus, the proposed closed-loop outer-loop dynamics admit an unforced dissipative port-Hamiltonian representation. The associated dissipativity and stability properties are established in Section III-E.

### E. Closed-Loop Dissipativity and Stability

For the state-dependent dissipation matrix $R(v)$ , it can be verified that when $\eta > 0$ , $k$>0, and $\rho > 0$ , $R \succ 0$ is positive definite. Differentiating H along the closed-loop system trajectories yields

$$\dot{H} = -\eta\left\|\nabla H\right\|^2 - \frac{k\rho}{\rho_h^*}\left[\hat{v}^T\nabla H\right]^2 \le 0 \tag{27}$$

For the admissible design parameters, the radial component of (24) is outward pointing in a punctured neighborhood of the origin; hence, $\mathcal{D}$ is forward invariant. The Hamiltonian function is monotonically non-increasing.

The stability proof of the closed-loop system is as follows. First, $H$ is radially unbounded: $H$ is a sum of three nonnegative squared terms. From (13), as $\|v\| \to \infty$ , the reactive power error $\tilde{Q}$ grows as $\rho^2$ , hence $H \ge \alpha\tilde{Q}^2/2$ tends to infinity as $\rho^4$ , so every sublevel set of $H$ is compact, and closed-loop trajectories are bounded. Second, there is no equilibrium point in the far field: when $\|v\|$ is sufficiently large, the dominant term in $\nabla H$ , namely $\alpha\tilde{Q}\,2v/\omega_0 L_g$ , points outward in the radial direction with magnitude $O(\rho^3)$ , while the remaining terms are at most $O(\rho^2)$ . Therefore, $\nabla H$ cannot vanish in the far field, and equilibrium points can only exist in a bounded region.

Under this premise, if the designed power-error weight guarantees that the desired equilibrium point $v^*$ is the unique stationary point of $H$ on $\mathcal{D} := \mathbb{R}^2 \setminus \{0\}$ , then LaSalle's invariance principle implies that every closed-loop trajectory in $\mathcal{D}$ converges to $v^*$ . It should be noted that, due to the polar-coordinate transformation and the $v/\rho$ term in the control law (24), the closed-loop vector field is undefined only at the origin. Therefore, the stability statement is understood relative to $\mathcal{D}$ . Thus, $v^*$ is globally asymptotically stable relative to $\mathcal{D}$ , or equivalently almost globally asymptotically stable in $\mathbb{R}^2$ .

Moreover, if there exists $m > 0$ such that $\nabla^2 H(v) \succeq mI$ locally around $v^*$ , then $H(v^*) = 0$ and $\|\nabla H(v)\|^2 \ge 2mH(v)$ . Combining this inequality with (27) gives $\dot{H} \le -2\eta mH$ , and hence $H(t) \le e^{-2\eta mt}H(0)$ . Therefore, the desired equilibrium is locally exponentially stable.

### F. Parameter Design

This section gives the selection range of the power-error weighting coefficient $\alpha$ , so that all undesired stationary points are excluded from the admissible domain $\mathcal{D} := \mathbb{R}^2 \setminus \{0\}$ . From (16), the stationary points satisfy $\nabla H = \tilde{\rho}\hat{v} + \alpha\tilde{P}\nabla P + \alpha\tilde{Q}\nabla Q = 0$ . Besides the expected high-voltage equilibrium point, if there exist other stationary points, then at points where $\tilde{P}\nabla P + \tilde{Q}\nabla Q \ne 0$ , the corresponding weight is

$$\alpha(v) = -\frac{\tilde{\rho}\hat{v}^T(\tilde{P}\nabla P + \tilde{Q}\nabla Q)}{\left\|\tilde{P}\nabla P + \tilde{Q}\nabla Q\right\|^2} \tag{28}$$

Substituting (23) into the stationary-point and the Hessian singularity condition and solving within $\mathcal{D} \setminus \{v_h^*\}$ yields the candidate bifurcation points. The real roots satisfying $\alpha(v) > 0$ are retained. At the bifurcation boundary, $\nabla^2 H_\alpha$ is positive semidefinite with one zero eigenvalue. Take the minimum value to define the critical power-error weight

$$\alpha_{cr} = \min_{v\in\mathcal{D}\setminus\{v_h^*\}} \alpha(v) \tag{29}$$

The above procedure guarantees that, for any $0 < \alpha < \alpha_{cr}$ , all undesired stationary points are excluded from $\mathcal{D}$ . Hence, the desired high-voltage equilibrium is the unique stationary point on $\mathcal{D}$ .

After the power-error weight is determined, the gradient dissipation gain $\eta$ and the radial dissipation gain $k$ are selected according to the desired convergence rate and bandwidth

$$\eta \geq \frac{\lambda_d}{m},\ 0 < k < \frac{\rho_h^*}{\rho_{\max}}(\frac{\omega_{\max}}{M} - \eta) \qquad (30)$$

where $\lambda_d$ is the desired decay rate of the state error; $m$ and $M$ are the uniform lower and upper bounds of the Hessian eigenvalues of the storage function in the certified operating domain, $\rho_{\max}$ denotes the maximum voltage magnitude within the certified operating region $\Omega_c$ and $\omega_{\max}$ is the allowable outer-loop bandwidth. The first inequality guarantees the prescribed local convergence rate, whereas the second limits the radial gain and enforces the prescribed outer-loop bandwidth.

### G. Energy Landscape Analysis

To intuitively illustrate the composition of the shaped Hamiltonian function in the last section, this section separately presents the energy landscapes of the Logistic amplitude potential energy, the power dispatch potential energy, and their superposition after shaping, revealing the geometric structure of the Hamiltonian function through energy landscape representation.

(1) Energy landscape of amplitude potential energy: Mapping the amplitude term in (13) to the two-dimensional voltage state space($v_x$, $v_y$), the energy landscape is shown in Fig. 1. The following structural features can be observed from Fig. 1: a valley floor is formed at $\rho = \rho_{\text{ref}}$, where the energy attains its global minimum of zero; the valley floor is rotationally symmetric in the phase angle direction, with zero energy gradient along the phase angle direction. This structure originates from the quadratic potential energy property of Logistic radial regulation: amplitude errors are driven to converge through the radial gradient.

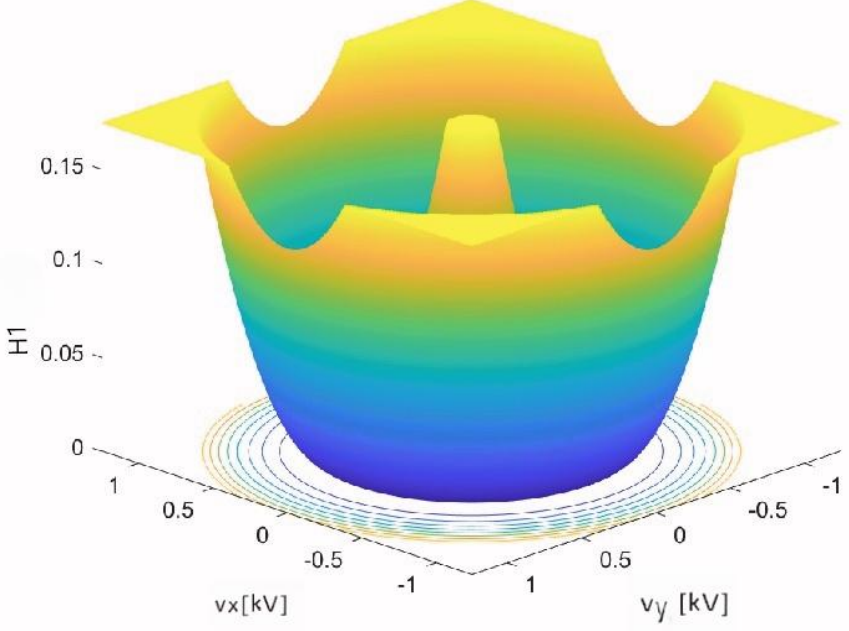

Fig. 1. Energy landscape of the Logistic amplitude potential energy.

(2) Energy landscape of $PQ$ power: The latter two terms in (13) constitute the power dispatch potential energy, as shown in Fig. 2. This power dispatch potential exhibits a double-valley landscape in the voltage state space. For a given power reference, two valley points exist in the state space, corresponding to the high-voltage and low-voltage steady-state power-flow solutions, respectively, separated by a saddle point. As can be seen from Fig. 2, when only power dispatch is used as the energy function, which valley point the system converges to depends on the basin of attraction into which the initial state falls, and convergence to the desired equilibrium point cannot be guaranteed. Therefore, the energy landscape needs to be reshaped through the intervention of the amplitude potential energy.

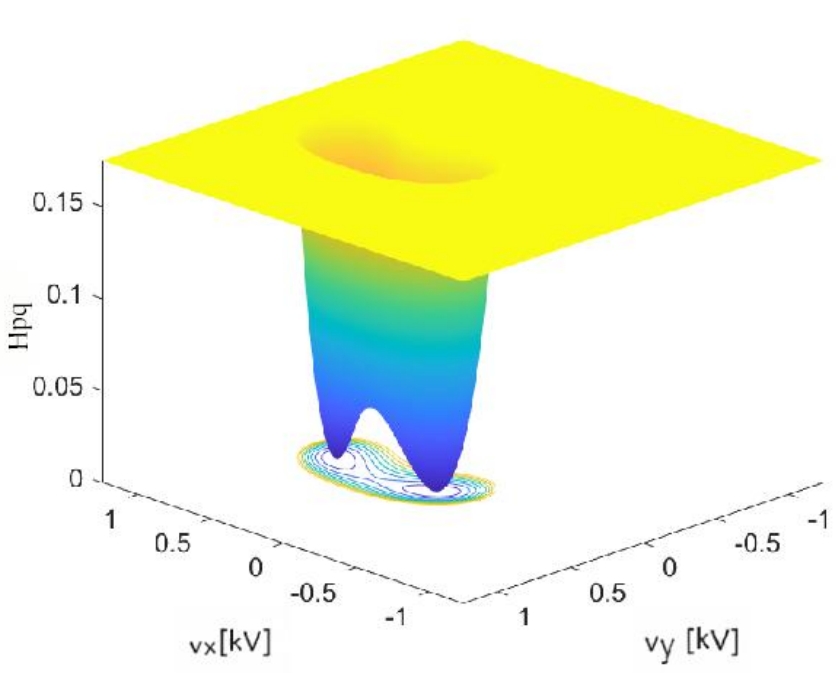

Fig. 2. PQ power energy landscape.

(3) Energy landscape after reshaping: Fig. 3 presents the energy landscape after shaping. Within the constructed certified operating region, the low-voltage power-flow solution no longer belongs to the set of closed-loop stationary points, rendering the desired high-voltage operating point the unique global minimum. On this basis, the subsequent control law design will take this energy landscape as the reference, driving the closed-loop dynamics to converge to this unique steady state along the energy descent direction.

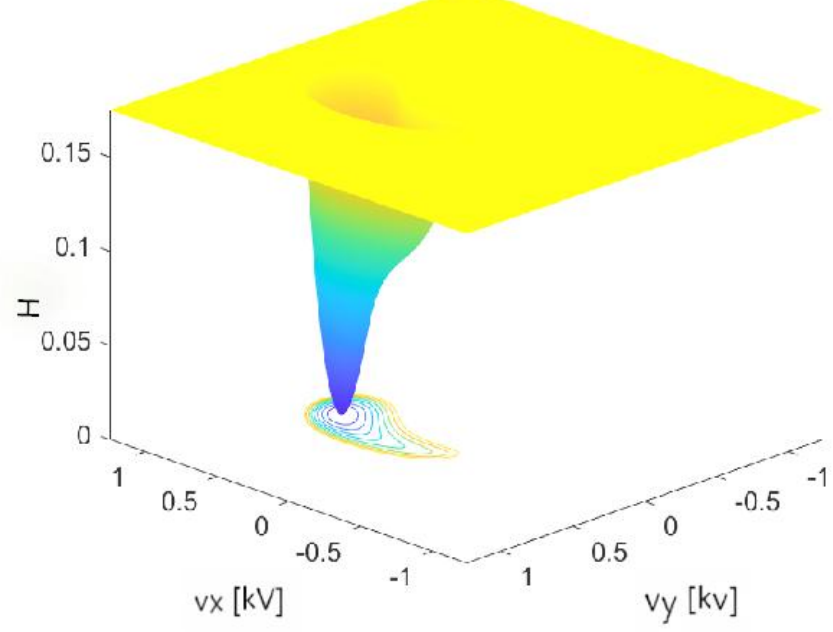

Fig. 3. Shaped energy landscape.

### H. Comparison with dVOC and PH-Based GFM Control

The proposed Logistic-VOC unifies amplitude regulation, synchronization, and power dispatch within a single shaped Hamiltonian, with the closed loop evolving as a dissipative port-Hamiltonian flow. Its radial channel coincides with the amplitude regulation of dVOC[2]; the distinction lies in the design logic. In dVOC, these functionalities enter as separate components, with almost-global stability established by a hierarchical, phase-first-magnitude-second analysis. Here, they arise from one geometric structure, stability follows from a direct energy-based argument, and undesired stationary points are excluded from the certified operating domain through potential-weight design. Likewise, unlike [15], where the PH structure relies on an auxiliary current loop and a switching pumping-or-damping action, the decomposition in (5) yields a direct dissipative PH form without either. Table I summarizes the conceptual comparison.

TABLE I COMPARISON WITH EXISTING DVOC AND PH-BASED GRID-FORMING CONTROL

| Aspect | dVOC [2] | PH-based design [15] | This paper |
|---|---|---|---|
| Energy function | $V(\bar{v}) = \bar{v}^T P \bar{v}$<br>Lyapunov function used for stability analysis | $H = H_P + H_C$,<br>$H_C = \frac{\Phi^2}{2\varepsilon^2}$, $\Phi = \frac{1}{2} C_f (v^2 - v_{ref}^2)$<br>Composite Hamiltonian | $H(v) = \frac{1}{2}\tilde{\rho}^2 + \frac{\alpha}{2}\tilde{P}^2 + \frac{\alpha}{2}\tilde{Q}^2$<br>Unified Hamiltonian |
| Design paradigm | Verification-based | Auxiliary-loop-based PH realization | Constructive energy-shaping design |
| Control law | $\dot{v} = \omega_0 J v + \eta_d (Kv - y) + \alpha_d (1 - \frac{\lVert v \rVert}{v^*}) v$<br>Not formulated in PH form | $\dot{x} = (J - R(x))\nabla(H_P + H_C) + L\xi, x = (v, i_L)^T$<br>Separate PH controller interconnected via an auxiliary current loop | $\dot{v} = (J - R(v))\nabla H(v)$<br>Direct dissipative PH form in the port-voltage state |
| Amplitude regulation | $\alpha_d (1 - \frac{\lVert v \rVert}{v^*}) v$<br>not explicitly expressed in gradient form | $C_f \xi_1 (v_{ref}^2 - v^2) v + C_f s_Q \xi_2 (\frac{Q_{ref}}{v_{ref}^2} - \frac{Q}{v^2}) v$ | $-\frac{k\rho}{\rho_{ref}} \nabla H_1(v), \rho = \lvert v \rvert$<br>$H_1 = \frac{1}{2}\tilde{\rho}^2$ |
| Dissipation & energy balance | No explicit $R$;<br>$\dot{V} < 0$ on $S^{\perp}$ under gain-network condition | $R_{sw} \succeq 0$ via switching $S_Q \in \{-1, 1\}$<br>$\dot{H} = \xi^T y - (\nabla H)^T R \nabla H \le \xi^T y$ | $R(v) = \eta I + \frac{k\rho}{\rho_h^*} \Pi_r$<br>$\dot{H} = -(\nabla H)^T R \nabla H \le 0$ |

## IV. SIMULATION VERIFICATION

This section verifies the energy shaping effectiveness of the proposed port-Hamiltonian Logistic-VOC grid-forming outer loop based on a grid-forming converter connected to the grid through a tie-line inductance. The simulation examines the energy landscapes of the shaped Hamiltonian function, closed-loop trajectories, and their time evolution under different power error weighting coefficients, verifying the constructed dissipation structure and high-voltage equilibrium point selection conditions.

In this case study, the system parameters are shown in Table II.

TABLE II SYSTEM SIMULATION PARAMETERS

| Rated power | $S_N = 6$ MVA |
|---|---|
| Grid line voltage (RMS) | $V_N = 0.69$ kV |
| Rated frequency | $f = 50$ Hz |
| Network inductance | $L_g = 0.2$ mH |
| Gradient dissipation gain | $\eta = 1.4$ |
| Radial dissipation gain | $k = 1.17$ |

For the given $P_{ref}$ = 2 MW and $Q_{ref}$ = 0.5 Mvar, the power-flow equation admits two solutions: $v_h = 0.687 + j0.182$ kV and $v_l = 0.002 + j0.182$ kV. The target point amplitude is $\rho_h^* = 0.711$ kV.

Based on the design method proposed in Section III-F, the critical power-error weight is calculated as $\alpha_{cr} = 0.0537$.

### A. High-Voltage Equilibrium Point Selection

Fig. 4 presents the energy contours and the convergence trajectories from typical initial points for the power error weighting coefficient at the supercritical value and the design value. Specifically, Fig. 4(a) corresponds to the supercritical parameter $\alpha$=0.06, where a low-voltage local minimum appears in the energy landscape, forming a saddle point and the corresponding energy separatrix between it and the high-voltage potential well. Among the selected initial points, the trajectory falling into the basin of attraction of the low-voltage potential well converges to the low-voltage local equilibrium, while the other trajectory converges to the desired high-voltage equilibrium point. Fig. 4(b) corresponds to the design parameter $\alpha$=0.04, in which case the shaped Hamiltonian admits the desired high-voltage equilibrium point as the unique stationary point. From different initial points, both state trajectories converge to the target point along the negative gradient direction of the energy function, numerically verifying the dissipativity and the unique-target attraction of the constructed closed loop.

These results confirm that $\alpha$ reshapes the energy landscape and hence the set of stationary points: once $\alpha$ exceeds $\alpha_{cr}$, the low-voltage attraction region reappears and convergence to the desired high-voltage equilibrium point is no longer guaranteed.

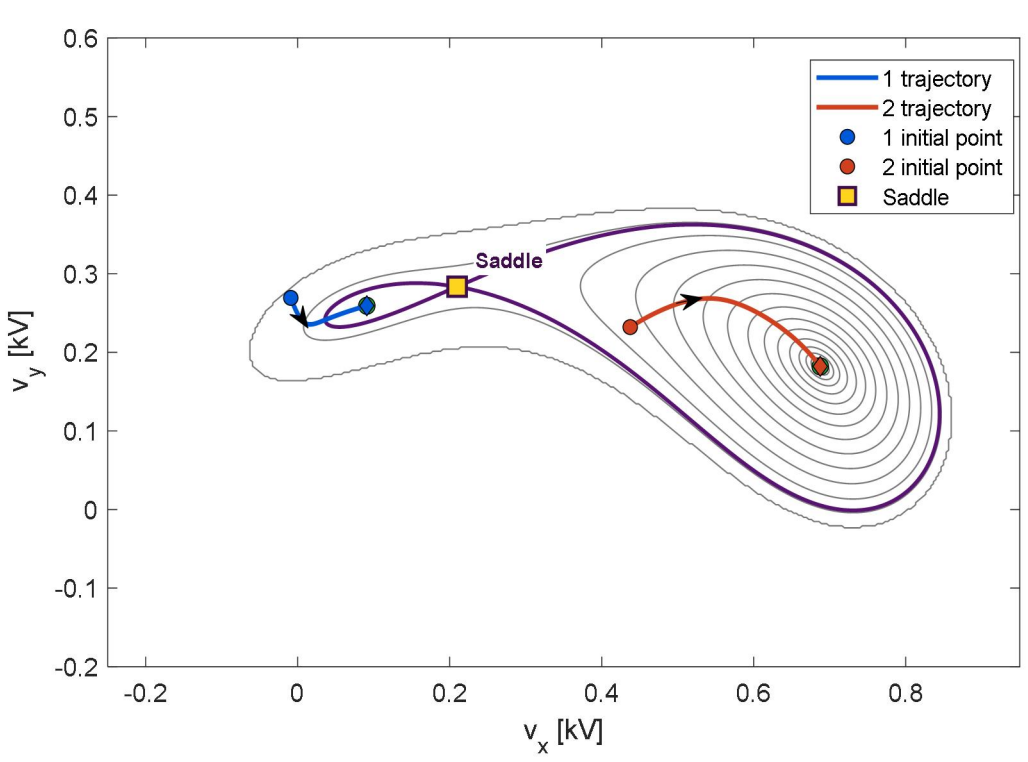

(a) Energy contours and initial point convergence trajectory for $\alpha$ = 0.06.

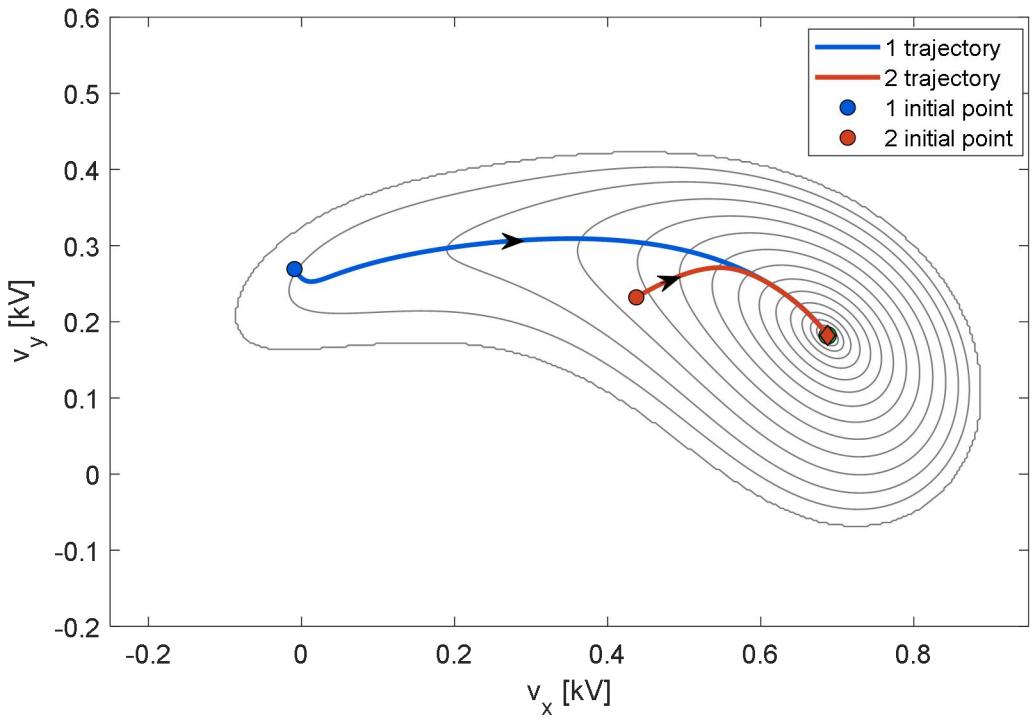

(b) Energy contours and initial point convergence trajectory for $\alpha$ = 0.04.

Fig. 4. Energy contours and convergence trajectories under different power error weighting coefficients.

### B. P/Q Dispatch-Step Response

To verify the $P/Q$ dispatch capability of the proposed PH-Logistic-VOC controller, two consecutive simultaneous active- and reactive-power reference steps are applied to the single grid-forming converter.

Before the disturbance, the converter operates at $P_{\text{ref}}$ = 2 MW and $Q_{\text{ref}}$ = 0.5 Mvar. At $t$ = 1 s, the references are simultaneously changed to $P_{\text{ref}}$ = 2.5 MW and $Q_{\text{ref}}$ = 0 Mvar, and at t = 2.5 s, both references are returned to their initial values.

Fig. 5 shows the active power $P$, reactive power $Q$, terminal-voltage magnitude|, and frequency deviation. Following the reference change, both active and reactive powers converge smoothly to their new set-points without sustained oscillation. The terminal-voltage magnitude evolves continuously from the pre-step high-voltage equilibrium point to the new high-voltage power-flow solution. Meanwhile, the frequency deviation remains bounded and rapidly decays to zero, with a peak value of approximately 0.15 Hz. These results verify that the proposed controller accomplishes simultaneous $P/Q$ dispatch while preserving stable convergence to the desired high-voltage operating branch.

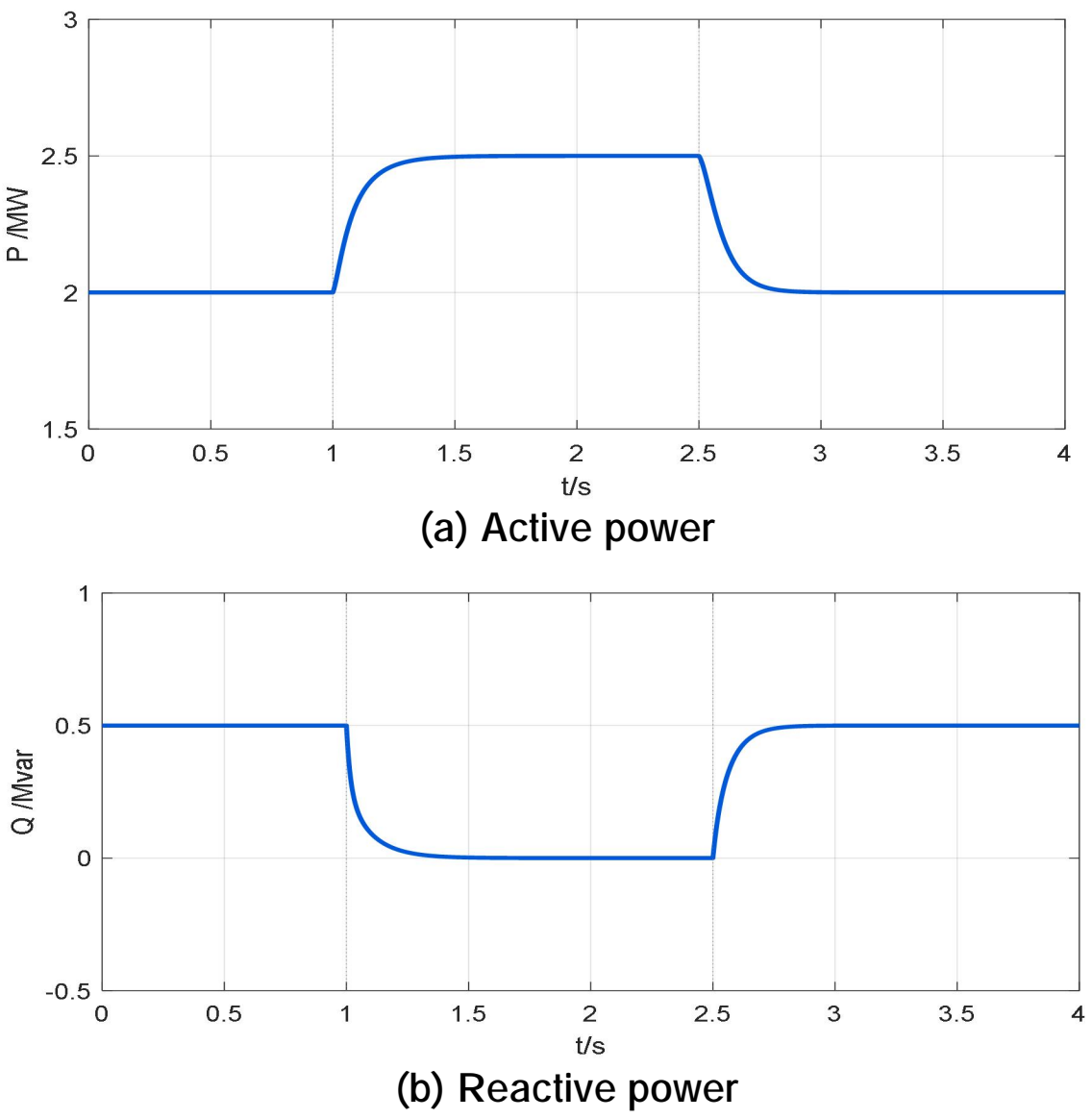

(a) Active power

(b) Reactive power

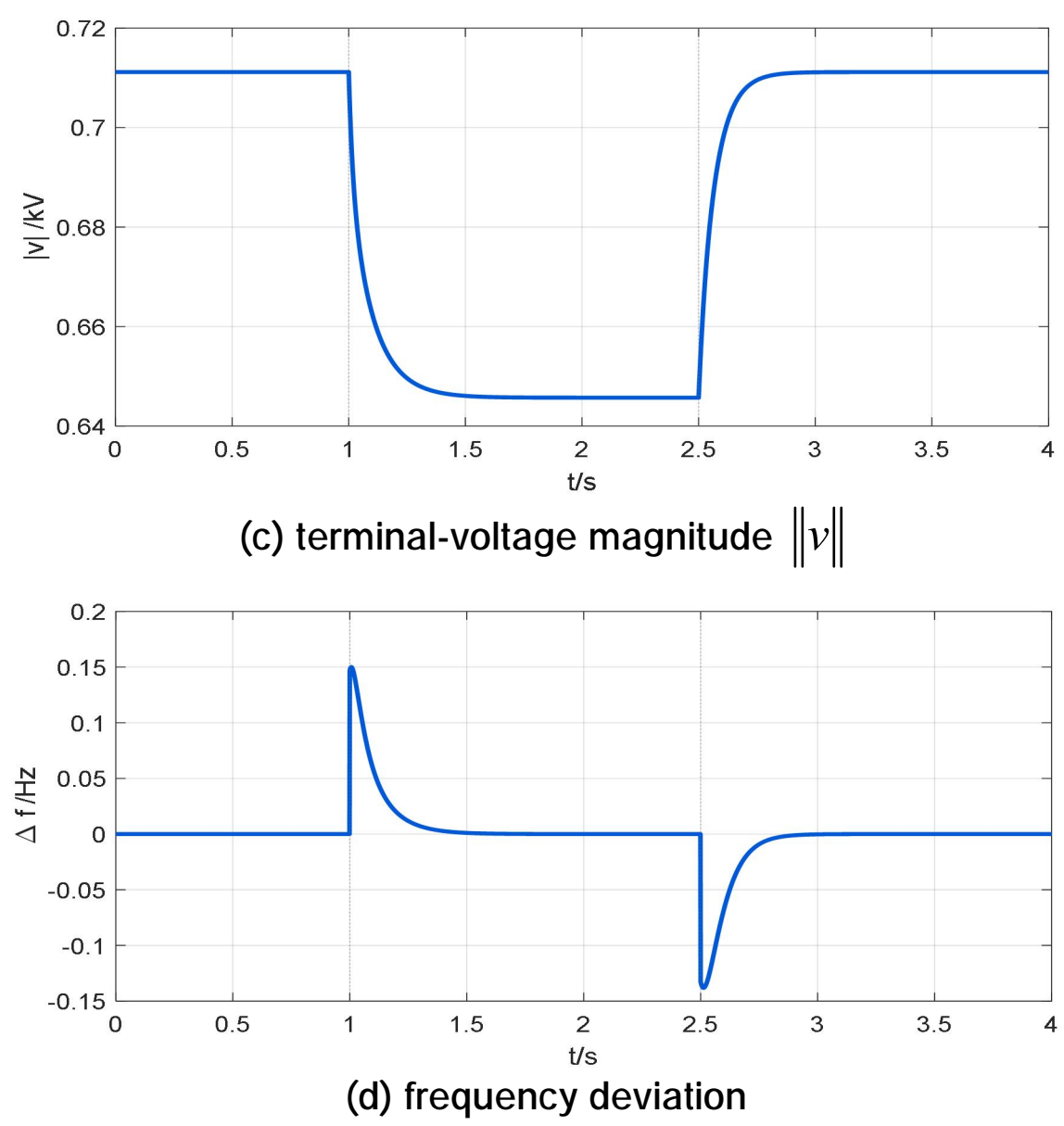

(c) terminal-voltage magnitude $\|v\|$

(d) frequency deviation

Fig. 5. Responses to P/Q dispatch-steps

### C. Transient Recovery under Large Disturbances in a Weak Grid

To assess the disturbance-rejection capability of the proposed PH-Logistic-VOC controller under weak-grid conditions, a scenario with a short-circuit ratio of SCR = 1.26 is considered. The active- and reactive-power references are maintained at 2.0 MW and 0.5 Mvar, respectively. Two representative large-disturbance cases are examined.

In Case A, a 0.8 pu grid-voltage sag lasting 200 ms is applied at $t$ =0.5 s. The corresponding transient responces are shown in Fig. 6.

In Case B, a 15° grid-voltage phase jump is imposed at $t$=0.5 s. The transient responses are presented in Fig. 7.

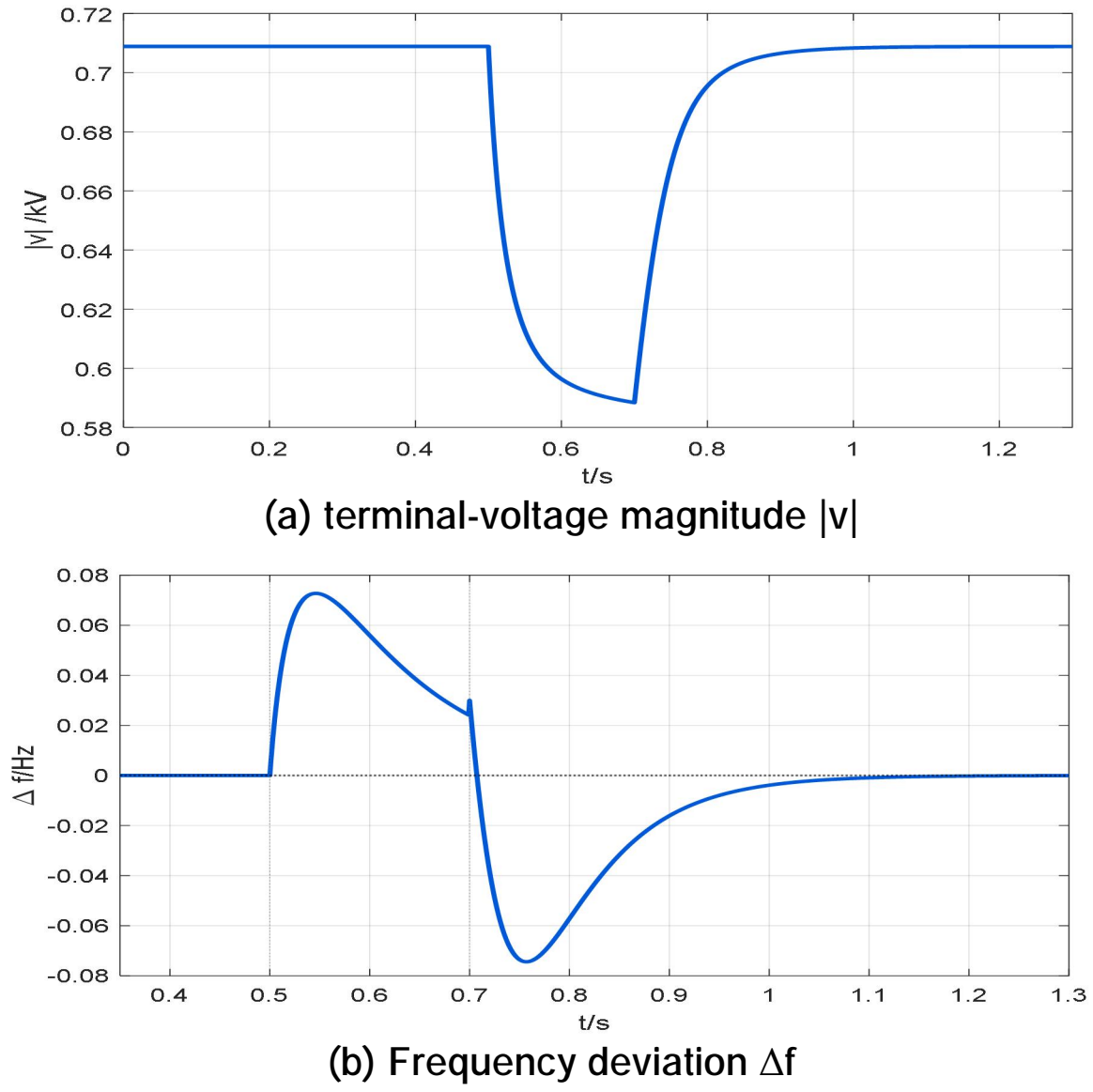

(a) terminal-voltage magnitude |v|

(b) Frequency deviation Δf

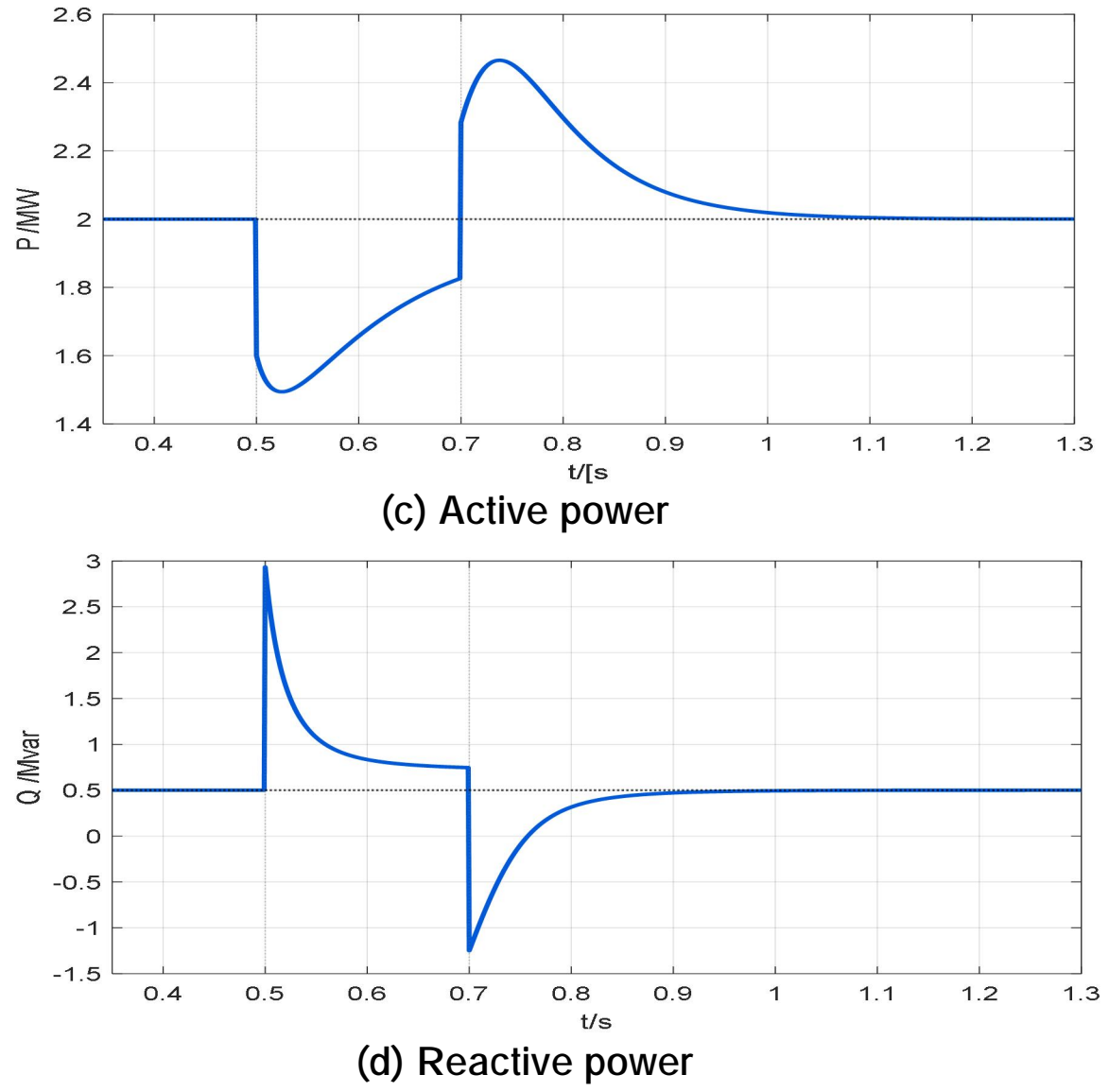

Fig. 6. Transient responses to a 0.8 pu grid-voltage sag at SCR = 1.26

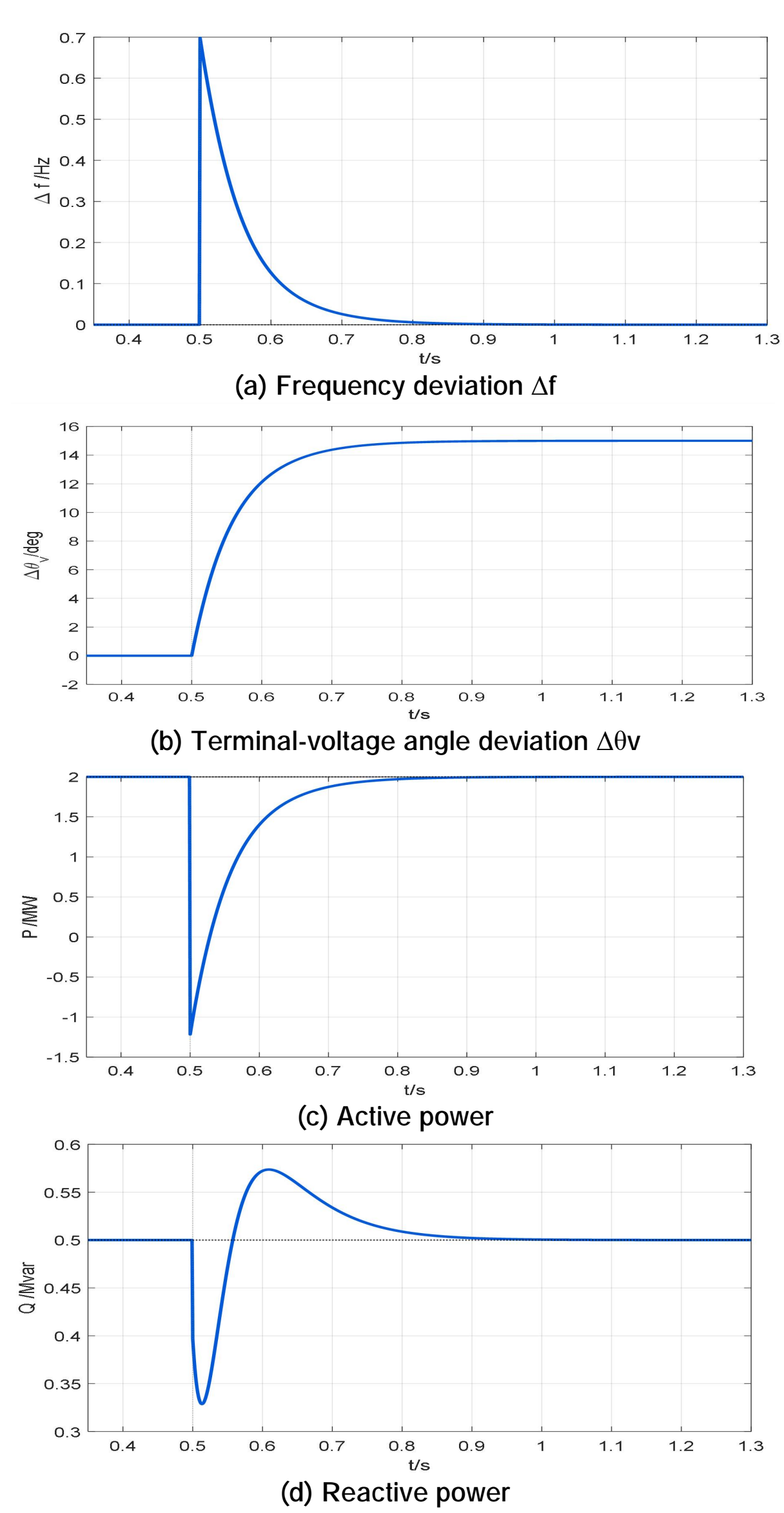

Fig. 7. Transient responses to a +15° grid-voltage phase jump at SCR = 1.26

As shown in Fig. 6, during the voltage sag, the terminal-voltage magnitude exhibits a bounded transient deviation and recovers after the disturbance is cleared. The active power, reactive power, and frequency deviation show only short-duration deviations without sustained oscillations. Similarly, as shown in Fig. 7, under the phase-jump disturbance, the terminal-voltage angle rapidly adjusts to the new grid-voltage angle, while the power and frequency signals return to their steady-state values after a brief transient. These results demonstrate that the proposed controller maintains stable operation under SCR = 1.26 weak-grid conditions and under large voltage-magnitude and phase disturbances, providing effective voltage recovery, power regulation, and disturbance attenuation.

## V. Conclusion

This paper proposes a constructive port-Hamiltonian energy-shaping design for Logistic-VOC-based grid-forming converters. A dissipative port-Hamiltonian outer-loop model is established with the port voltage vector as the state variable, revealing the compatibility between Logistic radial regulation and the quadratic energy storage structure, and unifying amplitude restoration and P/Q power dispatch as an energy shaping problem, thereby achieving gradient-based power dispatch control. Moreover, a state-dependent dissipation matrix is constructed to realize radial gain matching, ensuring that the closed-loop system strictly satisfies the port-dissipation inequality. Subsequently, almost-global asymptotic stability and local exponential convergence of the target operating point are proved. Furthermore, the regulation mechanism of the power error weighting coefficient on the set of closed-loop stationary points is revealed, and active selection of the desired high-voltage equilibrium point is achieved through energy landscape shaping. Simulation results validate the theoretical findings, demonstrating the effectiveness of the proposed method in power dispatch tracking and transient disturbance rejection. The proposed port-Hamiltonian outer-loop model provides a structured foundation for future extensions to inner-loop dynamics, current limiting, and multi-converter network coupling.

## References

[1] B. B. Johnson, M. Sinha, N. G. Ainsworth, F. Dörfler, and S. V. Dhople, "Synthesizing virtual oscillators to control islanded inverters," *IEEE Trans. Power Electron*., vol. 31, no. 8, pp. 6002–6015, Aug. 2016.

[2] M. Colombino, D. Groß, J.-S. Brouillon, and F. Dörfler, "Global phase and magnitude synchronization of coupled oscillators with application to the control of grid-forming power inverters," *IEEE Trans. Autom. Control*, vol. 64, no. 11, pp. 4496–4511, Nov. 2019.

[3] G.-S. Seo, M. Colombino, I. Subotić, B. Johnson, D. Groß, and F. Dörfler, "Dispatchable virtual oscillator control for decentralized inverter-dominated power systems: Analysis and experiments," *in Proc. IEEE Appl. Power Electron. Conf. Expo. (APEC)*, Anaheim, CA, USA, Mar. 2019, pp. 561–566.

[4] S. Azizi Aghdam and M. Agamy, "Small signal stability analysis of dispatchable virtual oscillator controlled inverters with adaptive virtual inertia control," *in Proc. IEEE Power Energy Soc. Innov. Smart Grid Technol. Conf. (ISGT)*, Washington, DC, USA, Jan. 2023, pp. 1–5.

[5] M. B. Abdelghany, M. Al Talaq, A. Al-Durra, and H. Zeineldin, "Nonlinear large signal stability analysis of dispatchable virtual oscillator control," *in Proc. IEEE Ind. Electron. Appl. Conf. (IEACon),* Kuala Lumpur, Malaysia, Nov. 2024, pp. 47–52.

[6] X. Shen and Y. Yang, "An improved dispatchable virtual oscillator control: Stability analysis and parameter design," *in Proc. IEEE Energy Convers. Congr. Expo. Europe (ECCE Europe)*, Birmingham, UK, Sep. 2025, pp. 1–6.

[7] M. Al Talaq, M. B. Abdelghany, A. Al-Durra, H. Zeineldin, and T. El-Fouly, "A sophisticated grid-forming dispatchable-virtual oscillator control with robust inner current control for improving transient stability," *IEEE Trans. Power Electron.*, vol. 40, no. 10, pp. 15064–15079, Oct. 2025.

[8] Z. Zeng, Y. Sun, and D. Yang, "Frequency-domain large-signal modeling and stability analysis for dispatchable virtual oscillator controlled grid-connected converters," *IEEE Open J. Power Electron.*, vol. 6, pp. 1068–1080, May 2025.

[9] D. Wang, "Port-Hamiltonian control of GFM-VSCs with robust stable and uniform error dynamics," *IEEE Access*, vol. 11, pp. 108752–108764, 2023.

[10] M. A. A. Murad, R. Ortega, J. I. Yuz, and S. V. Dhople, "Enhanced synchronization stability of grid-forming inverters with passivity-based virtual oscillator control," *IEEE Trans. Power Electron.*, vol. 37, no. 12, pp. 14141–14156, Dec. 2022.

[11] Y. Gui and Y. Xue, "Passivity-based control of grid forming and grid following converters in microgrids," *in Proc. IEEE Power Energy Soc. Gen. Meeting* (PESGM), Orlando, FL, USA, Jul. 2023, pp. 1–5.

[12] Y. Gui, S. Subedi, and Y. Xue, "Passivity-based grid forming control for DERs," *in Proc. IEEE Energy Convers. Congr. Expo. (ECCE)*, Phoenix, AZ, USA, Oct. 2024, pp. 3673–3677.

[13] M. Li, H. Geng, and X. Zhang, "Robust passivity-based control for grid-forming converter," *in Proc. IEEE Int. Electr. Energy Conf.* (CIEEC), Harbin, China, May 2023, pp. 2603–2608.

[14] M. Li, E. Liu, H. Geng, Y. Mao, X. Wang, and X. Zhang, "Passivity-based control for the stability of grid-forming multi-inverter power stations," *IEEE Trans. Ind. Electron.*, vol. 72, no. 9, pp. 9117–9127, Sep. 2025.

[15] L. Kong, Y. Xue, L. Qiao and F. Wang, "Control Design of Passive Grid-Forming Inverters in Port-Hamiltonian Framework," *IEEE Transactions on Power Electronics*, vol. 39, no. 1, pp. 332-345, Jan. 2024.